\documentclass[%
aip,
jmp,%
amsmath,amssymb,
preprint,%
]{revtex4-1}

\usepackage{graphicx}
\usepackage{dcolumn}
\usepackage{bm}
\usepackage{color}
\usepackage[mathlines]{lineno}

\begin{document}

\title[A Cryogenic Infrared Calibration Target]{A Cryogenic Infrared Calibration Target}
\author{E.J. Wollack}
\email{edward.j.wollack@nasa.gov}
\affiliation{NASA Goddard Space Flight Center, Greenbelt, MD 20771}
\author{R.E. Kinzer Jr.}
\affiliation{NASA Goddard Space Flight Center, Greenbelt, MD 20771}
\author{S.A. Rinehart}
\affiliation{NASA Goddard Space Flight Center, Greenbelt, MD 20771}
\date{\today}

\begin{abstract}
A compact cryogenic calibration target is presented that has a peak diffuse reflectance, $R \le 0.003$, from $800-4,800\,{\rm cm}^{-1}$ $(12-2\,\mu$m). Upon expanding the spectral range under consideration to $400-10,000\,{\rm cm}^{-1}$ $(25-1\,\mu$m) the observed performance gracefully degrades to $R \le 0.02$ at the band edges. In the implementation described, a high-thermal-conductivity metallic substrate is textured with a pyramidal tiling and subsequently coated with a thin lossy dielectric coating that enables high absorption and thermal uniformity across the target. The resulting target assembly is lightweight, has a low-geometric profile, and has survived repeated thermal cycling from room temperature to $\sim4\,$K. Basic design considerations, governing equations, and test data for realizing the structure described are provided. The optical properties of selected absorptive materials -- Acktar Fractal Black, Aeroglaze Z306, and Stycast 2850 FT epoxy loaded with stainless steel powder -- are characterized and presented. 
\end{abstract}


\keywords{Lossy Dielectric Coatings; Calibration Targets; Infrared Instrumentation and Techniques}

\maketitle

\begin{quotation}
\end{quotation}

\section{\label{sec:introduction}Introduction}
Absorptive targets and light traps find widespread use in flux calibration, termination of residual reflections in optical systems,  and establishing the zero point in reflectance spectrometry.~\cite{Palchetti2008} Approximating a near-ideal absorber in a finite volume presents a challenge for such applications -- the absence of reflected light or how ``black" an object appears is not a unique material property or treatable as a boundary condition within the framework of electromagnetic theory~\cite{Sommerfeld1950} -- but an object's reflectance and absorptance are intimately tied to the surface's material properties and the underlying geometry of the structure.~\cite{Hultst1957} Thus to implement a low reflectance calibrator design, the absorber material's dielectric and magnetic properties need to be either known or experimentally determined in order to suitably tailor the target's geometry. 

At microwave through sub-millimeter wavelengths, precision absorptive standards find use in radiometric flux calibration and low reflectance targets for remote sensing,~\cite{Gaidis1999} astrophysics,~\cite{Gush1992,Mather1999} and other metrology applications.~\cite{Janz1987} Elements of the design techniques used in these microwave structures can be applied to realize compact, light-weight, and low-reflectance absorbers while maintaining the overall manufacturability for far infrared use at cryogenic temperatures. In this work, a calibration target for cryogenic applications is explored and its performance is described in detail.  The design, manufacture and optical characterization of the target structure and selected coatings of potential interest are summarized in  Sections~\ref{sec:design}, \ref{sec:fabrication}, and \ref{sec:characterization} respectively.

\section{\label{sec:design}Target Design Considerations}
Low reflectance targets have been realized from individual geometric structures~\cite{Gush1992,Mather1999} or tilings~\cite{Fixsen2006} of absorbing dielectric cones or pyramids. Wedges, rulings~\cite{Gaidis1999,Janz1987,Kogut2004} and Brewster angle light traps~\cite{Elterman1977,Breneman1981,Clarke1986,LaRocca1985} have also been employed in instrumentation settings where only a single polarization state is of interest. Through the microwave waveband, the absorptive coating's finish is typically smooth compared to the freespace wavelength, $\lambda_o$, and the resulting reflections from surfaces making up the target's facets are largely specular. In this limit, the effective reflectance for a pyramidal wedge with vertex angle,  $\alpha$, can be estimated from geometric optics considerations,   $R \approx R_o^{\pi/\alpha}$.  Here the single bounce reflection at the angle of incidence, $R_o$, is estimated from the Fresnel coefficients.~\cite{Carli1974} In effect, the geometry can be used to limit the impact of imperfect {\it a priori} knowledge of the coating properties on the final target absorptance at cryogenic temperatures.~\cite{Wollack2007}  As a result, the key geometric parameters that control the reflectance in this limit are the details of the pyramid's points, grooves at the structure base, and the number of reflections within the structure.~\cite{Fixsen2006} Ideally these features would be perfectly sharp; however, in practice a finite length scale is achievable and set by fabrication tolerances and the coating morphology. 

In the limit where the freespace wavelength, $\lambda_o$, is less than twice the tiling period, $p$, the target's response can be computed by treating the structure as an artificial dielectric adiabatic impedance taper.~\cite{Meyer1956,Kuester1994} As the wavelength is further increased beyond the length scale of the taper,  $h$, the target's surface appears smooth and incident fields evanescently propagate into the structure.~\cite{Baekelandt1999} As a result, the radiation is largely specularly reflected from the structure.  More generally, these arguments can be extended to treat the diffuse reflectance from the surface roughness in the design process if sufficient details are available.~\cite{Bedford1974} Here the experimentally observed diffuse reflectance is taken as a property of the coating that in turn determines acceptable substrate geometry and the short wavelength limit of the structure's response. 

The use of an appropriate coating with a controlled and sufficient thickness to absorb the incident radiation is also an important consideration. Here the material is assumed to be non-magnetic and characterized as a function of frequency by a complex dielectric function, $\epsilon_r = \epsilon_r' + i \cdot \epsilon_r''$, relative to the permittivity of freespace. The thickness of the target's lossy dielectric layer employed must be greater than the material's field penetration length scale,  $\delta = \lambda_o/\pi\sqrt{2\epsilon_r''}$, to suitably attenuate signals before reflection from the underlying substrate.~\cite{Wollack2007} By design, the substrate is optically thick and the absorptance is related to the reflectance by, $A = 1 - R$. Consideration of the Fresnel coefficient magnitude at the surface of the lossy dielectric reveals that a lower index coating leads to lower insertion loss per bounce and thus has greater absorptance from the overall structure in the long wavelength limit. In the infrared, a reduction in coating thickness is desirable as long as the coating maintains structural integrity and minimal thermal gradients. 

With appropriate attention to detail a variety of coatings could potentially be employed in this application~\cite{Smith1984,Betts1985,Persky1999} to achieve the desired response.  The service temperature of epoxies, paints, and deposited films are typically limited by the material's adhesion and thermal match to the substrate material over the temperature range of interest. For high temperature applications, chemical breakdown and aging may also require more careful consideration. At cryogenic temperatures typical anodized coating (thermal conductivity, $\kappa \sim 1\,$W/m/K) with thickness, $\delta x <100\,\mu$m, produces a thermal gradient across the absorber, $\Delta T / q  \simeq \delta x/ \kappa \sim 0.1\,$K/mW per unit incident flux, $q$. For our application this consideration is subdominant to the $\sim0.2\,$K thermometry calibration uncertainty at 4\,K. Cooling or achieving precise knowledge of the target temperature becomes a greater concern in going from bulk to low-volume-filling-fraction metal or semiconducting foams~\cite{Advena1993,Becker1999} and semi-metallic nanotubes~\cite{Quijada2011} in the long wavelength limit due to such coatings' lower thermal conductivity. Additional considerations for space qualification and other low-outgassing applications can be found in the literature.~\cite{Kauder2006,Kralik2009,Salomon2009} Based upon prior experience for the target implementation described here we use a custom specified thickness of Acktar Fractal Black~\cite{Acktar} (Acktar FB) amenable to a variation on the vendor's standard ($11\,\mu$m thickness coating) process for the absorptive coating.

\begin{figure}
	\includegraphics[width=3.4 in]{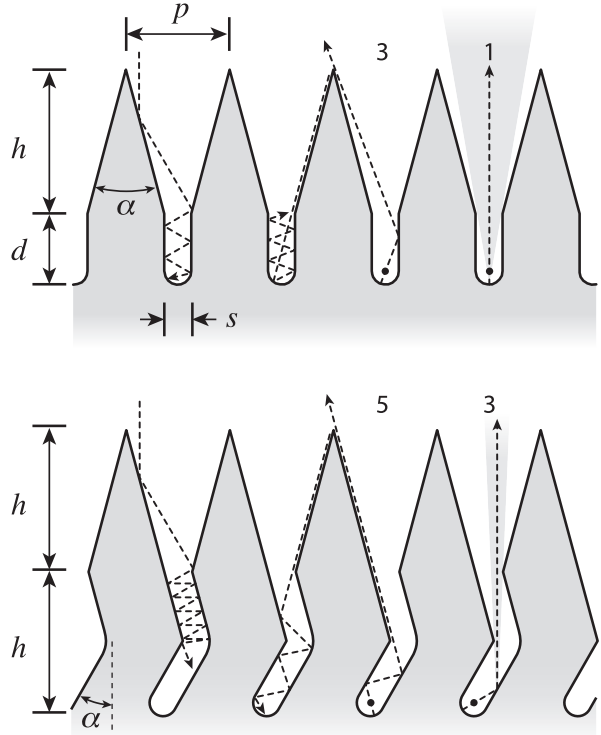}
	\caption{ The shaded gray regions in each figure represent the high thermal conductivity and optically thick patterned metallic substrate. The solid black lines are the lossy dielectric layer with thickness greater than the electrical penetration length, $\delta$. Representative geometric ray traces are indicated with dashed lines. The minimal number of bounces occur when rays go through the focus formed by the slot's radius (black dots, right).  For a simple slot, the single bounce observation angle is controlled by the slot width and depth (shaded region, upper right).  The overall absorptance can be improved through the addition of a ``dog leg" in the slot as shown (lower right).  }
	\label{Fig1_Disk_crossection}
\end{figure}

\section{\label{sec:fabrication} Sample Preparation and Fabrication}
In implementing the absorptive calibrator design we concentrate on the use of simple manufacturing approaches to achieve the desired response in the infrared. The underlying approach adopted was to use a lossy dielectric on a highly conductive substrate to achieve high absorptance. See Figure~\ref{Fig1_Disk_crossection} for target geometry and ray-trace summary. An aluminum alloy, 6061-T6, is chosen as the target's substrate material.  It is durable, easy to machine, and its high thermal conductance minimizes thermal gradients and enables knowledge of the device temperature. The design described is driven by a need for the target to have a total thickness $<6\,$mm. A total calibrator diameter, $d_{\rm cal} = 25\,$mm, was chosen for compatibility with our existing integrating sphere configuration and for use as a low-reflectance reference in a cryogenic filter wheel assembly. A pyramidal tiling of the surface is used to limit the total required thickness of the target relative to a single pyramid. 

The tips of the pyramids (radius $<20\,\mu$m) used in the tiling were readily achievable by wire electrostatic discharge milling (EDM) with a full apex angle, $\alpha = \pi/6$, and are comparable to the surface roughness features of the coatings employed. A finite width slot was substituted for the sharp groove found at the base of the pyramid in a traditional microwave target design. Slots $s \simeq 250\,\mu$m in width, $d = 940\,\mu$m deep, on a  pitch $p = 1370\,\mu$m were defined by the EDM wire diameter. The finite slot width relative to the pitch limits the achievable reduction in the pyramidal tiling thickness by a factor of $2 d_{\rm cal}  h / p^2 \sim 50$. The resulting patterned features in the aluminum substrate are less than half of the component's thickness and the remainder of the substrate is used to provide low lateral thermal conduction, mounting points for cooling, and thermometry. 

A cross-section of the fabricated pyramidal target's pyramidal tiles are shown in the upper panel of Figure~\ref{Fig1_Disk_crossection}. In the specular limit, this geometry has $>6$ reflections for $80\%$ of rays incident on the pyramidal structure. The ratio of the slot width over the depth limits number of rays with a single bounce and sets the target reflectance in this structure. By tipping the target to $\sim10^\circ$ from the normal the specular reflectance contribution from the bottom of the slot can be reduced. Alternatively, a ``dog leg" at the bottom of the channel as shown in the lower panel of Figure~\ref{Fig1_Disk_crossection} can be used to improve the absorptance by increasing the number of bounces ($N\ge3$) in the slot region and effectively reduces the angular extent of the caustic arising from the slot's finite radius. If greater volume for the target is available, the influence of the finite slot width can be further reduced by using an adiabatically curved path for the slot to geometrically hide its termination from view. 

\begin{figure}[!h]
	\includegraphics[width=3.4 in]{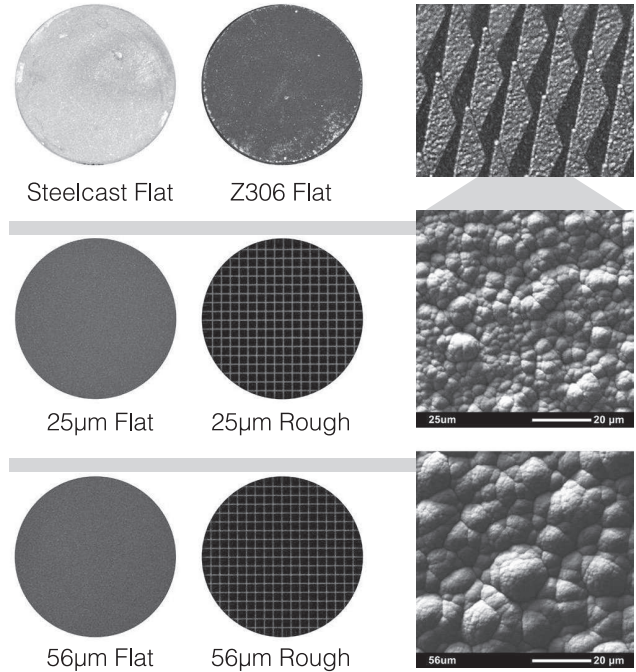}
	\vspace{-10pt}
	\caption{Images of substrate samples: 
(top-left) $300\,\mu$m Steelcast (Stycast 2850 FT epoxy loaded with $30\%$ stainless steel powder by volume) sample on aluminum flat; 
(top-center) $40\,\mu$m Aeroglaze Z306 black paint on Steelcast sample for enhanced long and short wavelength response; 
(top-right) Detail of Acktar FB coated calibrator on a substrate roughened with a pyramidal tiling viewed approximately $\pi/4$ from vertical; 
(middle-left) $25\,\mu$m Acktar FB witness coating on aluminum flat; 
(middle-center) $25\,\mu$m Acktar FB calibrator roughened with a pyramidal tiling; 
(middle-right) SEM image of $25\,\mu$m Acktar FB coating; 
(lower-left) $56\,\mu$m Acktar FB witness coating on aluminum flat; 
(lower-center) $56\,\mu$m Acktar FB calibrator roughened with a pyramidal tiling; and 
(lower-right) SEM image of $56\,\mu$m Acktar FB coating. 
The images of the ``flat" and ``rough" samples (i.e., left- and center-columns) were taken as a single exposure under an intense visible light source from approximately normal incidence. The logarithmic brightness scale has been uniformly adjusted in all digital images of the samples to convey the full range of the intensity recorded. The grey scale in the scanning electron microscope images (i.e., middle- and lower-right panels of figures) are uncorrected and convey the typical morphology of the Acktar FB samples investigated.}
	\label{Fig2_Disk_Photo_SEM}
\end{figure}

The pyramidally tiled absorber structures and optical witness flats were subsequently coated with Acktar FB~\cite{Acktar} via Physical Vapor Deposition (PVD) for spectral characterization. Two coating thicknesses, $\sim25\,\mu$m and $\sim50\,\mu$m, were specified in order to insure sufficient coating absorption. See Figure \ref{Fig2_Disk_Photo_SEM} for detailed digital images and a summary of the samples optically characterized in this study. 

%
%

In addition, a flat sample was coated with ``Steelcast'',  a dielectric mixture comprised of Emerson Cumming 2850 FT epoxy with a $30\%$ volume filing fraction of $8\,\mu$m stainless steel powder.~\cite{Wollack2008} The sample was then hand lapped to $\sim300\,\mu$m thickness with 600\,grit sand paper. The relative dielectric permittivity and magnetic permeability of this dielectric mixture are $\epsilon_r \approx 10.6+2.1i$ and $\mu_r \approx 1$, respectively. These properties were verified for the mixture employed by using a witness coupon as a waveguide Fabry-Perot resonator~\cite{Wollack2008} at $1\,{\rm cm^{-1}}$. To explore improving the Steelcast low frequency reflectance, a sample was coated with an anti-reflection layer realized from a $\sim40\,\mu$m thick layer of Aeroglaze Z306 paint~\cite{Aeroglaze} with an RMS surface roughness, $\sigma \simeq 2\,\mu$m. In modeling this multi-layer dielectric coating~\cite{Yeh1988,Smith1984} an effective relative permittivity, $\epsilon_r \approx 2.6+0.6i$,  is estimated for the paint on the epoxy substrate using the measured sample parameters and the specular reflectance $< 500\,{\rm cm^{-1}}$ (e.g., see the spectrum in Fig.~\ref{Fig3_Reflectance_Specular} labeled ``AG:Z306"). 

This combination of materials finds use in targets and baffles in cryogenic applications below $\sim400\,{\rm cm^{-1}}$ and provides an informative point of reference to the microwave structures motivating the calibrator design presented. Loaded epoxy structures are typically fabricated by building up the absorber in layers via spraying, molding, or with great difficultly direct machining on a textured substrate. The achievable tolerances with these approaches have higher variability than is observed with PVD coating on an identical patterned substrate in the design's critical tip and slot regions. As a result, a significantly greater calibrator volume is required to insure identical optical performance in practice and the structure's thermal time constant and gradients are correspondingly increased. While relatively high index loaded epoxies provide a path in the microwave to minimize the coating thickness and mass, in going to the infrared, lower index lossy thin films and conversion coatings more naturally serve this purpose.  Realizing the desired substrate geometry presents the primary challenge to be address in this path to calibrator volume reduction.

\section{\label{sec:characterization}FTS Characterization}
Infrared and sub-millimeter spectra were obtained using a Bruker IFS125hr Fourier transform spectrometer (FTS). This instrument is equipped with internal DTGS (Deuterated Triglycine Sulfate) detectors for use in the near- through far-infrared and an external LHe-cooled bolometer with two interchangeable filters for far-infrared and sub-millimeter wavelengths ($100-700$ and $5-100\,{\rm cm}^{-1}$, respectively). An array of beamsplitters is used for the different spectral regions; the following were used here: Si on CaF2 (near-IR), Ge on KBr (mid-IR), $6\,\mu$m multi-layer Mylar (Bruker T222/2), and $75\,\mu$m Mylar (far-infrared and sub-millimeter). The FTS is equipped with three internal light sources: tungsten (near-IR), SiC Globar (mid-IR), and a mercury arc lamp (far-IR). Reflectance spectra for the samples were measured in each band and then merged together to form a composite spectrum. Spectra are recorded {\it in vacuo} ($\sim0.3\,$mbar).

\begin{figure}[!h]
	\includegraphics[width=3.4 in]{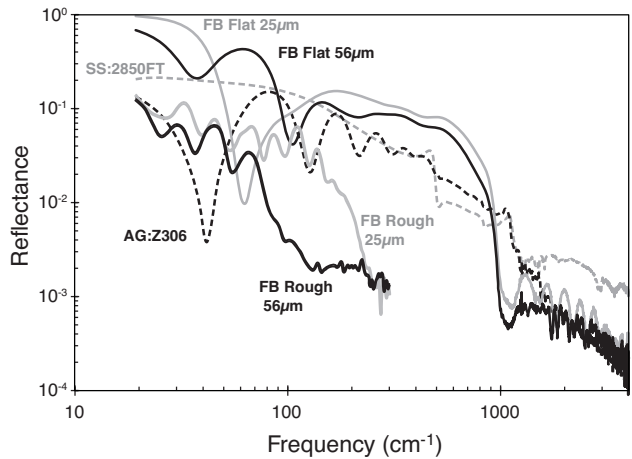}
	\caption{Measured specular reflectance. 
Dashed lines denote the stainless steel loaded 2850 FT epoxy (gray) and the Aeroglaze Z306 paint (black); 
solid lines are the Acktar FB flat $25\,\mu$m (gray) and $56\,\mu$m (black) coatings; 
and the bold lines are the Acktar FB $25\,\mu$m (gray) and $56\,\mu$m (black) coatings on substrates roughened with a pyramidal tiling. 
The coatings are deposited on aluminum substrates. }
	\label{Fig3_Reflectance_Specular}
\end{figure}

\begin{figure}[h]
	\includegraphics[width=3.4 in]{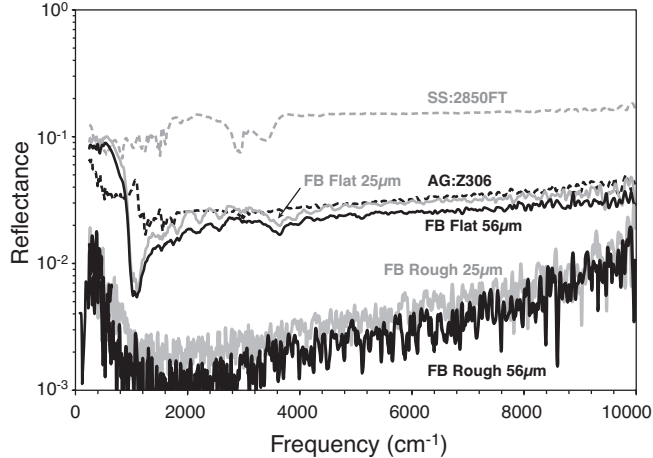}
	\caption{Measured diffuse reflectance. 
Dashed lines denote the stainless steel loaded 2850 FT epoxy (gray) and the Aeroglaze Z306 paint (black); 
solid lines are the Acktar FB flat $25\,\mu$m (gray) and $56\,\mu$m (black) coatings; 
and the bold lines are the Acktar FB $25\,\mu$m (gray) and $56\,\mu$m (black) coatings on substrates roughened with a pyramidal tiling. 
All coatings are deposited on aluminum substrate disks. The room temperature Bruker integrating sphere (300\,K, $>250\,{\rm cm}^{-1}$) and the OPASI-T cryogenic reflectometer (4.2\,K, $<650\,{\rm cm}^{-1}$) data were calibrated with a polished aluminum reference. The measured data from the two instrument configurations are in agreement in the region of shared spectral overlap.}
	\label{Fig4_Reflectance_Diffuse}
\end{figure}

Specular reflectance data were recorded over the frequency range of $20-5000\,{\rm cm}^{-1}$ at a $2\,{\rm cm}^{-1}$ resolution using a Pike Technologies accessory (Part No. SKU2084). Light from the spectrometer takes a “W”-path through the accessory, which has collimating optics and a fixed $10^\circ$ angle of incidence on the sample. In this set-up, the sample's reflectance spectrum is obtained by taking the ratio of the light intensity detected from the sample being measured to the intensity of an gold coated flat provided with the Pike accessory. See Figure \ref{Fig3_Reflectance_Specular} for the observed specular reflectance spectra.

Diffuse reflectance spectra were recorded using a Bruker integrating sphere accessory (Part No. A562G/Q) over a frequency range of $250-10000\,{\rm cm}^{-1}$. This accessory is equipped with a side-mounted DTGS detector used primarily for mid- and near- infrared wavelengths and is supplied with a diffusely reflecting gold reference. Internal absorptive baffles are used to shield the detector from receiving light reflected directly from the samples. In order to mitigate the typical noise of measurements with the integrating sphere apparatus, and to help enhance the broad structure of the features (e.g., see the Acktar FB $25\,\mu$m flat coating between $\sim 1000-3000\,{\rm cm}^{-1}$), spectra were recorded using spectral resolutions of 8 and $16\,{\rm cm}^{-1}$. In addition to the Bruker integrating sphere accessory, diffuse reflectance was measured for the roughened Acktar FB $56\,\mu$m coating using a custom cryogenic integrating sphere.~\cite{Rinehart2011} This apparatus is equipped with a Si bolometer element (4.2\,K) and an internal temperature-controlled sample wheel. Far infrared spectra between $100-650\,{\rm cm}^{-1}$ with $8\,{\rm cm}^{-1}$ resolution were measured using this apparatus mounted to a Bruker IFS 113v FTS equipped with an internal SiC Globar source and 3.5 and $12\,\mu$m Mylar beamsplitters. See Figure \ref{Fig4_Reflectance_Diffuse} for observed diffuse reflectance spectra.

\vspace{0pt}
\section{\label{sec:Discussion}Discussion}
From the observed diffuse reflectance of the witness samples in Figure~\ref{Fig4_Reflectance_Diffuse} one notes both Acktar FB coatings have a slightly lower reflection than Aeroglaze Z306 paint in the spectral range of interest. The calibrator constructed with $56\,\mu$m of Acktar FB was noted to have the best overall optical performance of the structures tested with an observed diffuse reflectance of $R<0.003$ between $800-4,800\,{\rm cm}^{-1}$ ($12-2\,\mu$m). The minimum in the diffuse reflectance,  $R\simeq 0.002$, occurs at $2000\,{\rm cm}^{-1}$ ($5\,\mu$m). This represents a reduction in the diffuse reflectance by a factor of $\sim 50$ over a simple flat with an identical coating. 

The low frequency response of the calibrator samples is set by the pyramidal tiling's pitch and finite coating thickness. At greater than $\sim10^\circ$ from normal incidence the grey lines seen in Figure~\ref{Fig2_Disk_Photo_SEM} that originate from spectral reflectance off the bottom of the calibrator's slots are no longer discernible. This artifact can be further mitigated by gradually turning the slot and removing its endpoint from view as indicated in Figure~\ref{Fig1_Disk_crossection}. Incorporation of this fabrication detail is recommended to improve the overall emittance. The calibrator is broadband, mechanically robust, and has survived multiple cool downs to $\sim4\,$K. Due to the structure's low reflectance it has been adopted as the preferred absorptive light trap for general laboratory use.

\vspace{0pt}
\section*{Acknowledgements}
The authors gratefully acknowledge financial support from the NASA ROSES/APRA program and thank E. Sharp and C. Wheeler for their contributions to preparation and fabrication of the structures presented here. R. Kinzer was supported by an appointment to the NASA Postdoctoral Program at GSFC, administered by the Oak Ridge Associated Universities.

\vspace{0pt}
%
%

\end{document}